# Transport coefficients in Lorentz plasmas with the power-law kappa-distribution

Du Jiulin[*]

*Department of Physics, School of Science, Tianjin University, Tianjin 300072, China*

Abstract

Transport coefficients in Lorentz plasma with the power-law $\kappa$-distribution are studied by means of using the transport equation and macroscopic laws of Lorentz plasma without magnetic field. Expressions of electric conductivity, thermoelectric coefficient and thermal conductivity for the power-law $\kappa$-distribution are accurately derived. It is shown that these transport coefficients are significantly modified by the $\kappa$-parameter, and in the limit of the parameter $\kappa \to \infty$ they are reduced to the standard forms for a Maxwellian distribution.

**I. INTRODUCTION**

Early in 1968, based on using observations of low-energy electrons in the magnetosphere, Vasyliunas analyzed energy spectra of observed electrons within the plasma sheet and drew an empirical function to model the velocity distribution of the electrons at high energies [1]. This velocity distribution was not a Maxwellian but a power-law with a parameter $\kappa$, and therefore it was later known as the $\kappa$-distribution. The $\kappa$-distribution reduced to a Maxwellian in the limit $\kappa \to \infty$. In fact, spacecraft measurements of plasma velocity distributions, both in the solar wind and in planetary magnetospheres and magnetosheaths, have revealed that non-Maxwellian distributions are quite common. In many situations the distributions appear reasonably Maxwellian at low energies but have a "suprathermal" power-law tail at high energies (see [2] and the references therein). This has been well modeled by the $\kappa$-distribution,

$$f_\kappa(\varepsilon) = B_k \left[ 1 + \frac{1}{\kappa} \frac{\varepsilon}{\varepsilon_0} \right]^{-(\kappa+1)}, \tag{1}$$

---
[*] E-mail address: jldu@tju.edu.cn

where $\varepsilon = \frac{1}{2}mv^2$ is kinetic energy of an electron with speed $v$ and mass $m$, $\varepsilon_0 = \frac{1}{2}mw_\kappa^2$ is a characteristic energy, with a characteristic speed $w_\kappa$ related to the most probable speed $w_0$ of a Maxwellian distribution with equal number and energy densities by $w_\kappa = w_0[(\kappa - \frac{3}{2})/\kappa]^{1/2}$ [2], and $B_\kappa$ is the normalization constant,

$$B_\kappa = \left[\pi w_0^2 (\kappa - \frac{3}{2})\right]^{-3/2} \frac{\Gamma(\kappa+1)}{\Gamma(\kappa-\frac{1}{2})}, \qquad (2)$$

where $w_0 \equiv (2k_B T/m)^{1/2}$ with temperature $T$ and Boltzmann constant $k_B$. Obviously, $\kappa > \frac{3}{2}$ and in the limit $\kappa \to \infty$ the $\kappa$-distribution (1) reduces to a Maxwellian one. In the solar corona, $\kappa$-like distributions have been proposed to arise from strong nonequilibrium thermodynamic gradients, Fermi acceleration at upwelling convection-zone waves or shocks, and electron-ion runaway in a Dreicerorder electric field (see [2] and the references therein, and recent refs.[3, 4] ).

In recent years, the researches on the plasmas with the $\kappa$-distribution as well as the $\kappa$-like power-law distributions have attracted great interest for their many interesting applications found in the wide fields of space plasma physics and astrophysics, and also for the $\kappa$-distribution family that can be studied under the framework of nonextensive statistics. These examples include the solar wind and suprathermal tails [5-8], nonextensivity in plasma [9], waves and instability [10-35], charging phenomena [36-41], and other related works [42-45] etc. For a long time, one has often been very curious to know about dynamical origin of the power-law distributions noted widely in physics, chemistry, biology and elsewhere. Most recently, by means of using the stochastic dynamics of Brownian motion one has found the condition under which power-law distributions of many different forms are generated in a nonequilibrium complex system [46, 47]. This condition is a generalized fluctuation-dissipation relation, i.e. an energy-dependent diffusion coefficient $D(\varepsilon)$ is associated with an energy-dependent friction coefficient $\gamma(\varepsilon)$ by the equation $D = m\gamma(k_B T + \kappa^{-1}\varepsilon)$. In the limit $\kappa \to \infty$ it becomes the standard form, $D=m\gamma k_B T$. Here it is useful to introduce the physical meaning of $\kappa$-parameter. If one makes a parameter replacement with $\kappa - \frac{3}{2} \equiv -(1-q)^{-1}$, then the $\kappa$-distribution (1) becomes a power-law $q$-distribution, $f_q(\varepsilon) \sim \left[1-(1-q)\varepsilon/k_B T\right]^{(5q-3)/2(1-q)}$, the form of which is similar to the $q$-distribution in nonextensive statistical mechanics [48]. In this way, according to the equation of $q$-parameter for nonequilibrium plasmas [9], the parameter $\kappa$ may be related to temperature gradient and Coulombian potential gradient, and therefore a deviation from infinity in parameter $\kappa \neq \infty$ measures a degree away from thermal equilibrium.

Plasma is composed of ions and electrons. If one assumes ion temperature to be equal to electron temperature, then speed of an electron is much faster than that of an ion because mass of an ion is much lager than that of an electron, so that one can believe that the ions appear to be stationary and the electrons are moving against the ions. Therefore, when calculating contribution of the electrons in plasma to the transport coefficients, one only needs to consider the electron-electron collisions and the electron-ion collisions. In plasma, if the electron-electron collisions can be neglected as compared with the electron-ion collisions, then the plasma is called Lorentz plasma. Lorentz plasma model is simple, but it is very suitable for studying some transport coefficients, such as electric conductivity etc, because the electron-electron collisions have no contribution to the conductivity.

Transport processes in nonequilibrium plasma involve a variety of thermodynamic "fluxes", such as heat flow and electric current etc. These "fluxes" are driven by the corresponding nonequilibrium thermodynamic gradients; e.g. heat flow is driven by temperature gradient, and electric current is driven by electric potential gradient. According to Onsager relation, the transport coefficients in Lorentz plasma without magnetic field appear in the following macroscopic laws [49]:

$$\mathbf{E} = \frac{1}{\sigma}\mathbf{j}_e + \alpha \nabla T, \tag{3}$$

$$\mathbf{j}_q = (\varphi + \alpha T)\mathbf{j}_e - \lambda \nabla T, \tag{4}$$

where $\mathbf{E} = -\nabla \varphi$ is electric field intensity, $\varphi$ is electric potential, $\sigma$ is electric conductivity, $\mathbf{j}_e$ is current density vector, $\alpha$ is thermoelectric coefficient, $\mathbf{j}_q$ is heat flux vector, and $\lambda$ is thermal conductivity in the case without electric current. In plasma, the distribution functions of particles are undoubtedly of great importance to the accurate calculations of transport coefficients. For the plasma following a Maxwellian distribution, one has known the expressions of transport coefficients $\sigma$, $\alpha$ and $\lambda$, but for the plasma following the power-law distribution and/or the $\kappa$-distribution, one has not known yet. In this work, I shall study the expressions of $\sigma$, $\alpha$ and $\lambda$ in the Lorentz plasma with the $\kappa$-distribution (1).

The paper is organized as follows. In sec. II, we introduce the transport equation for the Lorentz plasma with the $\kappa$-distribution. In sec. III, for the $\kappa$-distribution we derive the expressions of three transport coefficients, including electric conductivity, thermoelectric coefficient and thermal conductivity. Finally in sec. IV, we give the conclusion.

**II. TRANSPORT EQUATION**

If one lets $f(\mathbf{r}, \mathbf{v}, t)$ be a single-particle distribution function of the electron with velocity $\mathbf{v}$ and at position $\mathbf{r}$ and time $t$, then the stationary transport equation for the Lorentz plasma without magnetic field is

$$\mathbf{v} \cdot \frac{\partial f}{\partial \mathbf{r}} - \frac{e\mathbf{E}}{m} \cdot \frac{\partial f}{\partial \mathbf{v}} = C(f), \tag{5}$$

where $C(f)$ is the collision term. Furthermore, if one supposes $f(\mathbf{r}, \mathbf{v}, t) = f_0(\mathbf{r}, \mathbf{v}) + \delta f(\mathbf{r}, \mathbf{v}, t)$, where $f_0(\mathbf{r}, \mathbf{v})$ is a stationary nonequilibrium distribution and $\delta f(\mathbf{r}, \mathbf{v}, t)$ is a small variation about $f_0(\mathbf{r}, \mathbf{v})$, then the collision term of Lorentz plasma can be written [49] as

$$C(f) = -\nu_{ei}(\upsilon)\delta f. \tag{6}$$

And the electron-ion collision frequency $\nu_{ei}(\upsilon)$ is given by

$$\nu_{ei}(\upsilon) = \frac{4\pi z e^4 N_e L^{(ei)}}{m^2 \upsilon^3}, \tag{7}$$

where $N_e$ is total electron number, $\upsilon$ is speed of the electron, $ze$ is charge of the ion, and $L^{(ei)} \equiv \ln \Lambda^{(ei)}$ is the electron-ion scattering factor determined by the Debye length.

In this way, after the small terms on the left side is neglected, Eq.(5) becomes

$$\mathbf{v}\cdot\frac{\partial f_0}{\partial \mathbf{r}} - \frac{e\mathbf{E}}{m}\cdot\frac{\partial f_0}{\partial \mathbf{v}} = -\nu_{ei}(\upsilon)\delta f. \tag{8}$$

If the electrons in the plasma follow the power-law $\kappa$-distribution, one has $f_0 = f_\kappa$, and then the equation (8) becomes

$$\mathbf{v}\cdot\frac{\partial f_\kappa}{\partial \mathbf{r}} - \frac{e\mathbf{E}}{m}\cdot\frac{\partial f_\kappa}{\partial \mathbf{v}} = -\nu_{ei}(\upsilon)\delta f. \tag{9}$$

Substituting Eq.(1) into Eq.(9), we can obtain

$$\delta f = -\frac{(1+\kappa)f_\kappa}{\nu_{ei}(\upsilon)[k_B T(\kappa-\tfrac{3}{2})+\varepsilon]}\left[e\mathbf{E} + \frac{\varepsilon}{T}\nabla T\right]\cdot\mathbf{v}. \tag{10}$$

## III. TRANSPORT COEFFICIENTS

### 1. The electric conductivity

In order to study the electric conductivity $\sigma$ in the plasma with the power-law $\kappa$-distribution, we can let $\nabla T = 0$ in Eq.(10), which has no effect on the calculation of $\sigma$. And thus Eq.(10) becomes

$$\delta f = -\frac{(1+\kappa)f_\kappa}{\nu_{ei}(\upsilon)[k_B T(\kappa-\tfrac{3}{2})+\varepsilon]}e\mathbf{E}\cdot\mathbf{v}. \tag{11}$$

By a velocity-space integral, the current density vector is defined as

$$\mathbf{j}_e = -eN_e\int\mathbf{v}f d\mathbf{v} = -eN_e\int\mathbf{v}\delta f\, d\mathbf{v}, \tag{12}$$

where we have determined that $\int \mathbf{v} f_\kappa d\mathbf{v} = 0$ because the integrand is an odd function of the velocity. Substituting Eq.(11) into Eq.(12), we have

$$\mathbf{j}_e = (1+\kappa)N_e e^2 \mathbf{E}\cdot\int\frac{\mathbf{v}\mathbf{v}}{\nu_{ei}(\upsilon)}\left[k_B T(\kappa-\tfrac{3}{2})+\varepsilon\right]^{-1} f_\kappa d\mathbf{v}$$

$$= \frac{1}{3}(1+\kappa)N_e e^2 \left\langle\frac{\upsilon^2}{\nu_{ei}(\upsilon)}\left[k_B T(\kappa-\tfrac{3}{2})+\varepsilon\right]^{-1}\right\rangle\mathbf{E}, \tag{13}$$

where we have determined that $\int \mathbf{v}\mathbf{v} F(\upsilon)d\mathbf{v} = \tfrac{1}{3}\mathbf{U}\int\upsilon^2 F(\upsilon)d\mathbf{v}$ with unit tensor $\mathbf{U}$, and introduced an average value defined by $\langle F(\upsilon)\rangle \equiv \int F(\upsilon)f_\kappa d\mathbf{v}$ (the same hereinafter). Comparing Eq.(13) with Eq.(3), in the case of $\nabla T = 0$ we find the electric conductivity,

$$\sigma = \frac{1}{3}(1+\kappa)N_e e^2 \left\langle \frac{\upsilon^2}{\nu_{ei}(\upsilon)}\left[k_B T(\kappa-\tfrac{3}{2})+\varepsilon\right]^{-1}\right\rangle. \tag{14}$$

Substituting Eq.(7) and $\varepsilon = m\upsilon^2/2$ into Eq.(14), we can write

$$\sigma = \frac{m^2(1+\kappa)}{12\pi k_B Tze^2 L^{(ei)}(\kappa-\tfrac{3}{2})}\left\langle \upsilon^5(1+A_\kappa \upsilon^2)^{-1}\right\rangle, \tag{15}$$

where we have used an abbreviation $A_\kappa \equiv m/2k_B T(\kappa-\tfrac{3}{2})$ (the same hereinafter). Calculating the average value in Eq.(15) (see Appendix), we obtain the expression of electric conductivity in the plasma following the $\kappa$-distribution,

$$\sigma = \frac{(\kappa-\tfrac{3}{2})^{3/2}\Gamma(\kappa+1)}{\kappa(\kappa-1)(\kappa-2)\Gamma(\kappa-\tfrac{1}{2})}\frac{4\sqrt{2}(k_B T)^{3/2}}{\pi^{3/2}m^{1/2}ze^2 L^{(ei)}}. \tag{16}$$

Clearly, new electric conductivity depends significantly on the $\kappa$-parameter. It is easily proved that in the limit $\kappa \to \infty$ the expression Eq.(16) can be reduced to the standard form of $\sigma$ for the plasma following a Maxwellian distribution [49],

$$\sigma = \frac{4\sqrt{2}(k_B T)^{3/2}}{\pi^{3/2}m^{1/2}ze^2 L^{(ei)}}. \tag{17}$$

## 2. The thermoelectric coefficient

In order to study the thermoelectric coefficient $\alpha$ in the plasma with the power-law $\kappa$-distribution, one can let **E**=0 in Eq.(10), which has no effect on the calculation of $\alpha$. And then Eq.(10) is written as

$$\delta f = -\frac{(1+\kappa)\varepsilon f_\kappa}{\nu_{ei}(\upsilon)T[k_B T(\kappa-\tfrac{3}{2})+\varepsilon]}\mathbf{\upsilon}\cdot\nabla T. \tag{18}$$

Substituting Eq.(18) into Eq.(12), the current density vector becomes

$$\mathbf{j}_e = (1+\kappa)eN_e T^{-1}(\nabla T)\cdot\int\frac{\mathbf{\upsilon\upsilon}}{\nu_{ei}(\upsilon)}\left[k_B T(\kappa-\tfrac{3}{2})+\varepsilon\right]^{-1}\varepsilon f_\kappa d\mathbf{\upsilon}$$

$$= \frac{(1+\kappa)eN_e}{3T}\left\langle\frac{\upsilon^2\varepsilon}{\nu_{ei}(\upsilon)}\left[k_B T(\kappa-\tfrac{3}{2})+\varepsilon\right]^{-1}\right\rangle\nabla T. \tag{19}$$

Comparing Eq.(19) with Eq.(3) and then using Eq.(7), in the case of **E**=0 we can find the thermoelectric coefficient,

$$\alpha = -\frac{(1+\kappa)N_e e}{3\sigma T}\left\langle\frac{\upsilon^2\varepsilon}{\nu_{ei}(\upsilon)}\left[k_B T(\kappa-\tfrac{3}{2})+\varepsilon\right]^{-1}\right\rangle$$

$$= -\frac{(1+\kappa)m^3}{24\pi z e^3 L^{(ei)} \sigma k_B T^2 (\kappa - \frac{3}{2})} \left\langle \upsilon^7 \left[1 + A_\kappa \upsilon^2\right]^{-1} \right\rangle. \tag{20}$$

Substituting the expression of $\sigma$, Eq.(15), into Eq.(20), we have

$$\alpha = -\frac{m}{2eT} \left\langle \upsilon^7 \left[1 + A_\kappa \upsilon^2\right]^{-1} \right\rangle \Big/ \left\langle \upsilon^5 \left[1 + A_\kappa \upsilon^2\right]^{-1} \right\rangle. \tag{21}$$

After these two averages in Eq.(21) are calculated respectively (see Appendix), we can obtain the expression of thermoelectric coefficient in the plasma following the power-law $\kappa$-distribution,

$$\alpha = -\frac{4k_B(\kappa - \frac{3}{2})}{e(\kappa - 3)}. \tag{22}$$

It is clear that new expression of thermoelectric coefficient depends significantly on the $\kappa$-parameter, and in the limit $\kappa \to \infty$ it can be reduced to the standard form for the plasma following a Maxwellian distribution [49], $\alpha = -4k_B / e$.

## 3. The thermal conductivity

In order to study the expression of thermal conductivity $\lambda$ in the plasma with the power-law $\kappa$-distribution, in the case without the electric current one can let $\mathbf{j}_e=0$ in Eq.(3) and employs $\mathbf{E} = \alpha \nabla T$, which has no effect on the calculation of $\lambda$. And thus Eq.(10) is written as

$$\delta f = -\frac{(1+\kappa)(\alpha e + \varepsilon/T) f_\kappa}{\nu_{ei}(\upsilon)[k_B T(\kappa - \frac{3}{2}) + \varepsilon]} \mathbf{v} \cdot \nabla T. \tag{23}$$

By a velocity-space integral, the heat flux vector is defined as $\mathbf{j}_q = N_e \int \mathbf{v} \varepsilon \, \delta f \, d\mathbf{v}$. Using Eq.(23) we have

$$\mathbf{j}_q = -N_e(1+\kappa)(\nabla T) \cdot \int \mathbf{vv} \frac{(\alpha e + \varepsilon/T)\varepsilon f_\kappa}{\nu_{ei}(\upsilon)[k_B T(\kappa - \frac{3}{2}) + \varepsilon]} d\mathbf{v}$$

$$= -\frac{1}{3} N_e (1+\kappa) \left\langle \frac{\upsilon^2 (\alpha e + \varepsilon/T)\varepsilon}{\nu_{ei}(\upsilon)[k_B T(\kappa - \frac{3}{2}) + \varepsilon]} \right\rangle \nabla T. \tag{24}$$

Comparing Eq.(24) with Eq.(4) and then using Eq.(7), in the case of $\mathbf{j}_e=0$ we can find the thermal conductivity,

$$\lambda = \frac{1}{3} N_e (1+\kappa) \left\langle \frac{\upsilon^2 (\alpha e + \varepsilon/T)\varepsilon}{\nu_{ei}(\upsilon)[k_B T(\kappa - \frac{3}{2}) + \varepsilon]} \right\rangle$$

$$= \frac{m^3(1+\kappa)}{24\pi z e^4 L^{(ei)} k_B T(\kappa - \frac{3}{2})} \left[ \frac{m}{2T} \left\langle \upsilon^9 (1+A_\kappa \upsilon^2)^{-1} \right\rangle + \alpha e \left\langle \upsilon^7 (1+A_\kappa \upsilon^2)^{-1} \right\rangle \right]. \qquad (25)$$

Calculating these two averages in Eq.(25) respectively (see Appendix) and then substituting Eq.(22) into it, we can obtain the expression of $\lambda$ in the plasma following the power-law $\kappa$-distribution,

$$\lambda = (\kappa - \tfrac{3}{2})^{\frac{7}{2}} \frac{\kappa+1}{\kappa-3} \frac{(\kappa-5)!}{\kappa!} \frac{\Gamma(\kappa+1)}{\Gamma(\kappa-\tfrac{1}{2})} \frac{16\sqrt{2} k_B (k_B T)^{5/2}}{\pi^{3/2} m^{1/2} z e^4 L^{(ei)}}. \qquad (26)$$

Clearly, new thermal conductivity depends significantly on the $\kappa$-parameter. It is easily proved that in the limit $\kappa \to \infty$ the expression Eq.(26) can be reduced to the standard form of $\lambda$ for the plasma following a Maxwellian distribution [49],

$$\lambda = \frac{16\sqrt{2} k_B (k_B T)^{5/2}}{\pi^{3/2} m^{1/2} z e^4 L^{(ei)}}. \qquad (27)$$

### IV. CONCLUSION

The power-law $\kappa$-distribution has attracted great interest for its many interesting applications found in the wide fields of space plasma physics and astrophysics, and also for the $\kappa$-distribution family that has been under the framework of nonextensive statistics. In conclusion, by means of using the transport equation and macroscopic laws of Lorentz plasma without magnetic field, we have studied three transport coefficients in the plasma with the $\kappa$-distribution, including electric conductivity, thermoelectric coefficient and thermal conductivity. We have accurately derived expressions of these three transport coefficients in the Lorentz plasma following the $\kappa$-distribution. New obtained expressions of the transport coefficients, i.e. $\sigma$, $\alpha$ and $\lambda$, are given by Eq.(16), Eq.(22) and Eq.(26), respectively. We have showed that these transport coefficients are modified significantly by the $\kappa$-parameter, and in the limit of the parameter $\kappa \to \infty$ they can reduce to the standard forms for the plasma following a Maxwellian distribution.

The present work is based on the "phenomenological" $\kappa$-distribution (1) with a characteristic speed: $w_\kappa = w_0[(\kappa - \frac{3}{2})/\kappa]^{1/2}$, where there are unphysical points for $\kappa \leq 3/2$. Of course, we can employ other $\kappa$-distributions such as that used by Vasyliunas [1] as the basic power-law distribution function to avoid the unphysical points. However, usually the parameter $\kappa$ is very large, and so the points have actually no effect on the practical applications.

**Acknowledgement**

This work is supported by the National Natural Science Foundation of China under Grants No. 11175128, No.10675088 and also by the Higher School Specialized Research Fund for Doctoral Program under Grant No. 20110032110058.

**Appendix: Integral calculations**

$$\begin{aligned}\langle \upsilon^5(1+A_\kappa\upsilon^2)^{-1}\rangle &= 4\pi \int_0^\infty \upsilon^7(1+A_\kappa\upsilon^2)^{-1} f_\kappa d\upsilon \\ &= 4\pi B_\kappa \int_0^\infty \upsilon^7(1+A_\kappa\upsilon^2)^{-(\kappa+2)} d\upsilon \\ &= \frac{12\pi B_\kappa}{(\kappa-2)A_\kappa} \int_0^\infty \upsilon^5(1+A_\kappa\upsilon^2)^{-(\kappa+2)} d\upsilon \\ &= \frac{24\pi B_\kappa}{(\kappa-1)(\kappa-2)A_\kappa^2} \int_0^\infty \upsilon^3(1+A_\kappa\upsilon^2)^{-(\kappa+2)} d\upsilon \\ &= \frac{24\pi B_\kappa}{\kappa(\kappa-1)(\kappa-2)A_\kappa^3} \int_0^\infty \upsilon(1+A_\kappa\upsilon^2)^{-(\kappa+2)} d\upsilon \\ &= 12\pi \frac{(\kappa-3)!}{(\kappa+1)!} \frac{B_\kappa}{A_\kappa^4}. \end{aligned} \quad (A1)$$

$$\begin{aligned}\langle \upsilon^7(1+A_\kappa\upsilon^2)^{-1}\rangle &= 4\pi \int_0^\infty \upsilon^9(1+A_\kappa\upsilon^2)^{-1} f_\kappa d\upsilon \\ &= 4\pi B_\kappa \int_0^\infty \upsilon^9(1+A_\kappa\upsilon^2)^{-(\kappa+2)} d\upsilon \\ &= \frac{16\pi B_\kappa}{(\kappa-3)A_\kappa} \int_0^\infty \upsilon^7(1+A_\kappa\upsilon^2)^{-(\kappa+2)} d\upsilon \\ &= 48\pi \frac{(\kappa-4)!}{(\kappa+1)!} \frac{B_\kappa}{A_\kappa^5}. \end{aligned} \quad (A2)$$

$$\begin{aligned}\langle \upsilon^9(1+A_\kappa\upsilon^2)^{-1}\rangle &= 4\pi \int_0^\infty \upsilon^{11}(1+A_\kappa\upsilon^2)^{-1} f_\kappa d\upsilon \\ &= 4\pi B_\kappa \int_0^\infty \upsilon^{11}(1+A_\kappa\upsilon^2)^{-(\kappa+2)} d\upsilon \\ &= \frac{20\pi B_\kappa}{(\kappa-4)A_\kappa} \int_0^\infty \upsilon^9(1+A_\kappa\upsilon^2)^{-(\kappa+2)} d\upsilon \\ &= 240\pi \frac{(\kappa-5)!}{(\kappa+1)!} \frac{B_\kappa}{A_\kappa^6}. \end{aligned} \quad (A3)$$